\documentstyle[preprint,aps,tighten,epsf,floats]{revtex}
\newcommand{\nn}{\nonumber\\}
\newcommand{\la}{\langle}
\newcommand{\ra}{\rangle}
\newcommand{\pslash}{\not\hspace{-0.7mm}p}
\newcommand{\ben}{\begin{displaymath}}
\newcommand{\een}{\end{displaymath}}
\newcommand{\be}{\begin{equation}}
\newcommand{\ee}{\end{equation}}
\newcommand{\bea}{\begin{eqnarray}}
\newcommand{\eea}{\end{eqnarray}}
\newcommand{\eqn}[1]{\label{#1}}
\newcommand{\eq}[1]{Eq.~(\ref{#1})}
\newcommand{\eqs}[1]{Eqs.\ (\ref{#1})}
\newcommand{\fign}[1]{\label{#1}}
\newcommand{\fig}[1]{Fig.\ \ref{#1}}

\newcommand{\bPhi}{\bar{\Phi}}
\newcommand{\bfp}{{\bf p}}

\newcommand{\bfq}{{\bf q}}

\begin{document}
\draft
\title{Gauging the spectator equations}
\author{A. N. Kvinikhidze\footnote{On leave from Mathematical Institute of
Georgian Academy of Sciences, Tbilisi, Georgia.} and B. Blankleider}
\address{Department of Physics, The Flinders University of South Australia,
Bedford Park, SA 5042, Australia}
\date{\today}
\maketitle
\begin{abstract}
We show how to derive relativistic, unitary, gauge invariant, and charge
conserving three-dimensional scattering equations for a system of hadrons
interacting with an electromagnetic field. In the method proposed, the spectator
equations describing the strong interactions of the hadrons are gauged using our
recently introduced gauging of equations method. A key ingredient in our model
is the on-mass-shell particle propagator. We discuss how to gauge this
on-mass-shell propagator so that both the Ward-Takahashi and Ward identities are
satisfied. We then demonstrate our gauging procedure by deriving the
gauge-invariant three-dimensional expression for the deuteron
photodisintegration amplitude within the spectator approach.
\end{abstract}


\pacs{21.45.+v, 24.10.Jv, 25.20.-x, 25.30.Bf, 25.30.Fj}

\section{Introduction}
Recently we have shown how to describe the interaction of an electromagnetic
probe with a hadronic system described by four-dimensional integral equations
\cite{G4d}.  Our method is based on the idea of gauging the integral equations
themselves, and in this way incorporates the electromagnetic interaction into
the hadronic description without the need for any perturbation expansion. As a
result, the external photon becomes attached to all possible places in every
contributing Feynman diagram of the theory, so that gauge invariance and charge
conservation are implemented in the theoretically correct fashion. In
Ref.~\cite{G4d} we applied the gauging of equations method to the
four-dimensional three-nucleon problem thereby obtaining gauge invariant
expressions for the electromagnetic currents of all possible transitions between
three-nucleon states induced by an external electromagnetic field. The power of
the method was particularly evident in the formulation of the three-body bound
state current where a previously overlooked overcounting problem was solved
automatically by the natural appearance of a subtraction term.

Combined with the integral equations describing the strong interactions, the
gauging of equations method provides a consistent unified description of
hadronic systems and their interactions with an external electromagnetic field.
Since the starting point of Ref.~\cite{G4d} was relativistic quantum field
theory, at this stage such a unified description is inherently
four-dimensional. In view of the technical difficulty in solving
four-dimensional equations, the question naturally arises if there is a way to
do a three-dimensional reduction of the unified description so that covariance,
unitarity, gauge invariance, and charge conservation are all preserved. This
paper is devoted to answering this question.

In the strong interaction sector, three-dimensional reductions of the
Bethe-Salpeter (BS) equation have been developed over a number of years
\cite{Log,Sug,Kad,Gross} and now provide a powerful approach for practical
calculations in quantum field theory. All these reductions preserve covariance
and unitarity, and in this respect give rise to the question of which reduction
is to be preferred \cite{Jack,Erk,Thom}. In Refs.~\cite{Gross}, Gross showed
that his reduction scheme has the important property of giving a
three-dimensional two-body equation that approaches the correct one-body
equation in the limit when one of the masses becomes very large. We find that
the Gross reduction scheme is also appealing in that it easily lends itself to
our gauging of equations method.

In the Gross approach, also called the ``spectator approach'', three-dimensional
equations are derived by restricting some of the intermediate state particles
(typically the spectator particles) in the BS equation to their mass
shell. Equivalently, the Feynman propagators $d$ of these particles in the BS
equation are replaced by the quantities $\delta$ containing a positive energy
on-mass-shell $\delta$-function:
\be
d(p)=\frac{i\Lambda(p)}{p^2-m^2+i\epsilon} \hspace{5mm}\rightarrow\hspace{5mm}
\delta(p)= 2\pi \Lambda(p) \delta^+(p^2-m^2)     \eqn{delta}
\ee
where $\Lambda(p)=1$ or ${\pslash+m}$ for scalar and spinor particles
respectively.  We shall call $\delta(p)$ the ``on-mass-shell particle
propagator''.  Thus in the two-body case, the propagator $G_0(P,p)=d_1(P-p)
d_2(p)$ in the BS equation
\be
T(P;k',k)=K(P;k',k)+\int \frac{d^4p}{(2\pi)^4} K(P;k',p)G_0(P,p)T(P;p,k)
\eqn{BS-eq}
\ee
is replaced by ${\cal G}_0(P,p)= d_1(P-p)\delta_2(p)$:
\be
{\cal G}_0(P,p)= \left \{ \begin{array}{ll}
2\pi d_1(P-p)\delta^+ (p^2-m_2^2) & \hspace{3mm}
\mbox{for scalar particle 2}\\[3mm]
2\pi d_1(P-p)\delta^+ (p^2-m_2^2)(\pslash +m_2) &\hspace{3mm}
\mbox{for spinor particle 2}.
\end{array}
\right.
\ee
This replacement turns the BS equation into the four-dimensional equation
\be
T(P;k',k)=K(P;k',k)+\int \frac{d^4p}{(2\pi)^4} K(P;k',p){\cal G}_0(P,p)T(P;p,k),
\eqn{Gross-eq}
\ee
which after a trivial integration over $p_0$ becomes the three-dimensional
``spectator equation'' for the t-matrix (in this sense, we shall also refer to
four-dimensional equations like \eq{Gross-eq} as being ``three-dimensional'').
The significance of expressing the three-dimensional spectator equation in the
four-dimensional form of \eq{Gross-eq}, is that we can then apply our gauging of
equations method directly to \eq{Gross-eq} in just the same way as was done for
the BS case in Ref.~\cite{G4d}.

Yet an immediate problem arises. As the gauging of an equation involves the
gauging of all terms in the equation, we are faced with having to gauge the
on-mass-shell one-body propagator $\delta(p)$ in \eq{Gross-eq}. The resulting
gauged on-mass-shell one-body propagator $\delta^\mu(p',p)$ needs to satisfy
both the Ward-Takahashi identity and the Ward identity if the overall gauging
procedure is to yield results that are gauge invariant and that obey charge
conservation (as we shall see later, it is possible for a gauged on-mass-shell
propagator to satisfy the Ward-Takahashi identity but not the Ward
identity). How to gauge $\delta(p)$ so that both these identities are satisfied
is therefore the key question that needs to be answered before a unified
three-dimensional description can be given. The major part of this paper is
devoted to answering this question. With this achieved, we then go on and
demonstrate the gauging procedure by deriving the gauge invariant
three-dimensional expression for deuteron photodisintegration within the
spectator approach. Application to the three-dimensional three-nucleon problem
is given in a separate work \cite{nnn3d}. Clearly, the gauging method we propose
is directly applicable to any system of hadrons for which the strong interaction
spectator equations can be written down.

It is also important to realise that although we concentrate our efforts in this
paper on the electromagnetic interaction for which gauge invariance (or current
conservation) is a major issue, the gauging of equations method itself is
totally independent of the type of external field involved. Thus the procedure
for obtaining three-dimensional equations for transition currents outlined in
this paper is valid ``as is'' for the case of other interactions (e.g. weak)
when the external field is that of a W or other gauge boson. Only gauged inputs
like the nucleon vertex function $\Gamma^\mu$ would need to be changed.

\section{Gauged on-mass-shell propagator}

To discuss the gauging of the one-body on-mass-shell propagator $\delta(p)$, it
is sufficient to consider a bound two-body system and its interaction with an
external electromagnetic field. In the BS approach, two-body scattering is
described by \eq{BS-eq} and the two-body bound state is described by the
equation
\be
\Phi_P(k)=\int \frac{d^4p}{(2\pi)^4}K(P;k,p)G_0(P,p)\Phi_P(p) \eqn{Phi-BS}
\ee
where $\Phi_P$ is the bound-state vertex function. Interaction with an external
electromagnetic field is then described by the bound-state current \cite{Mand}
\bea
\lefteqn{\langle P'|J^{\mu}(0)|P\rangle =
\int \frac{d^4p}{(2\pi)^4}\bar{\Phi}_{P'}(p')d_1(P-p)d_2^{\mu}(p',p)
\Phi_{P}(p)}\hspace{1cm}\nn
&&+ \int \frac{d^4p}{(2\pi)^4} \bar{\Phi}_{P'}(P-p)d_1^{\mu}(p',p)
d_2(P-p)\Phi_{P}(P-p)\nn
&&+\int \frac{d^4k}{(2\pi)^4}\frac{d^4p}{(2\pi)^4}\bar{\Phi}_{P'}(k)
d_1(P'-k)d_2(k)K^{\mu}(P',k;P,p)d_1(P-p)
d_2(p)\Phi_{P}(p)        \eqn{Mand} 
\eea
where $q=P'-P=p'-p$ is the four-momentum of the incoming photon, $K^{\mu}$ is
the interaction current, and
\be
d_i^\mu(p',p) = d_i(p')\Gamma_i^\mu(p',p)d_i(p)    \eqn{d^mu}
\ee
is the gauged Feynman propagator for particle $i$, with $\Gamma_i^{\mu}(p',p)$
being the particle's electromagnetic vertex function. \eq{Mand} is illustrated
in \fig{dgamma}. A simple way to derive \eq{Mand} is to gauge the BS equation
for the two-body Green function \cite{G4d}.  As $d_i$ is the propagator of a
particle without dressing, consistency requires that $\Gamma_i^{\mu}(p',p)$ be
the bare electromagnetic vertex, i.e.  for a scalar or spinor particle of charge
$e_i$, $\Gamma_i^{\mu}(p',p)= e_i(p'+p)^{\mu}$ or $e_i\gamma^{\mu}$,
respectively. The case where dressing is included does not add to the essential
discussion of this paper and is therefore relegated to the Appendix.
\begin{figure}[t]
\hspace*{3cm}  \epsfxsize=11cm\epsfbox{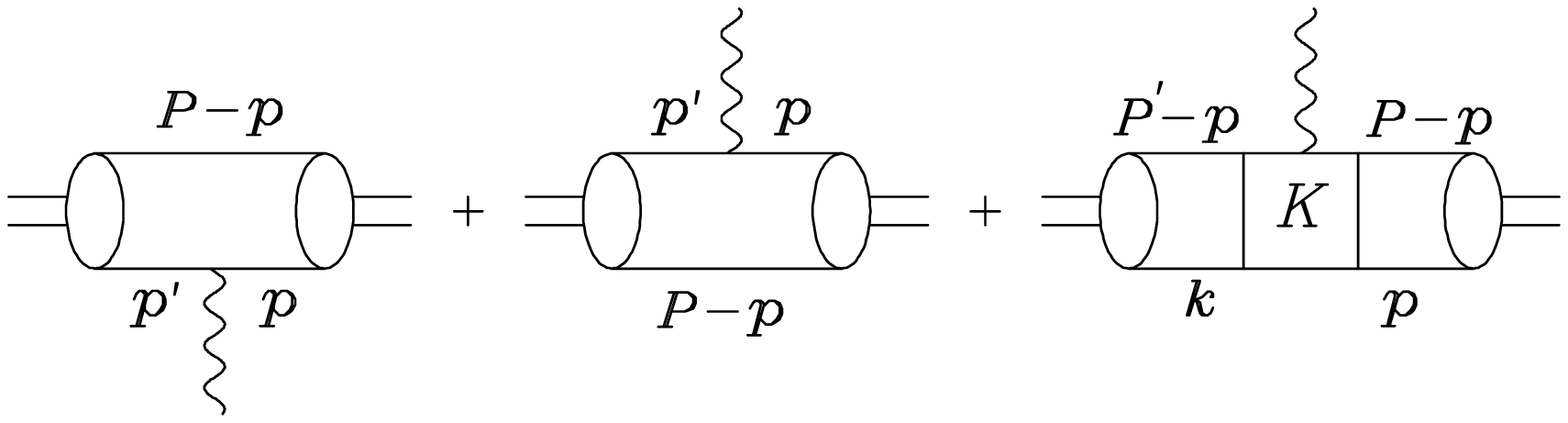}
\caption{\fign{dgamma} The two-body bound state current $\la
P'|J^\mu(0)|P\ra$ as given by \protect\eq{Mand}.}
\end{figure}

The three-dimensional reduction of \eqs{Phi-BS} and (\ref{Mand}) by putting
particle 2 on mass shell was discussed by Gross and Riska (GR)
\cite{GR}. Replacing $d_2$ by $\delta_2$ in \eq{Phi-BS} gives the bound state
spectator equation
\be
\Phi_P(k)=\int \frac{d^4p}{(2\pi)^4}K(P;k,p){\cal G}_0(P,p)\Phi_P(p).
\eqn{Phi-Gross}
\ee
In \eq{Mand} $d_2$ can be replaced by $\delta_2$ in the second and third terms
on the r.h.s.\ of the equation (2nd and 3rd terms of \fig{dgamma}), thus
reducing the four-dimensional integrations to three-dimensional ones, and at the
same time reducing the BS bound state vertex functions to the quasipotential
ones. Unfortunately it is impossible to do the same replacement for both
propagators of $d_2^\mu(p',p) = d_2(p')\Gamma_2^\mu(p',p)d_2(p)$ in the first
term on the r.h.s.\ of \eq{Mand} (1st term of \fig{dgamma}), as at the very
least this would make the bound state current diverge at zero momentum
transfer. To avoid this problem, GR replaced the first
term by a sum of two terms corresponding to particle 2 being on mass shell
to the right and to the left of the photon. That is, their prescription
is equivalent to the following gauge invariant replacement
\be
d^{\mu}(p',p) \hspace{3mm}\rightarrow\hspace{3mm}  \delta'^{\mu} (p',p)
= \delta(p')\Gamma^{\mu}(p',p)d(p)
+ d(p')\Gamma^{\mu}(p',p)\delta(p).      \eqn{dGR}
\ee
Although this prescription has been used in a number of calculations
\cite{Van1,Van2,Sur}, we shall see below that it leads to the breaking of charge
conservation.  For this reason, here we propose a different gauge invariant
replacement
\be
d^{\mu}(p',p) \hspace{3mm}\rightarrow\hspace{3mm} \delta^{\mu}(p',p)= i \frac
{\delta(p')\Gamma^{\mu}(p',p)\Lambda(p)-
\Lambda(p')\Gamma^{\mu}(p',p)\delta(p)}{p^2-p'^2}    \eqn{delta^mu}
\ee
which does lead to charge conservation. \eq{delta^mu} can also be written in
the form
\be
\delta^{\mu}(p',p)=2\pi i\Lambda(p')\Gamma^{\mu}(p',p)\Lambda(p)\frac
{\delta^+(p'^2-m^2)-\delta^+(p^2-m^2)}{p^2-p'^2} ,  \eqn{new}
\ee
showing that $\delta^{\mu}(p',p)$ is explicitly regular at $p^2-p'^{2}=0$. Using
this replacement, together with that of \eq{delta}, the bound state current of
\eq{Mand} is reduced to the three-dimensional expression
\bea
\lefteqn{\langle P'|J^{\mu}(0)|P\rangle =
\int \frac{d^4p}{(2\pi)^4}\bar{\Phi}_{P'}(p')d_1(P-p)\delta_2^{\mu}(p',p)
\Phi_{P}(p)} \hspace{1cm}\nn
&&
+ \int \frac{d^4p}{(2\pi)^4} \bar{\Phi}_{P'}(P-p)d_1^{\mu}(p',p)
\delta_2(P-p)\Phi_{P}(P-p)  \eqn{Mand_x}      \\ 
&&
+\int \frac{d^4k}{(2\pi)^4}\frac{d^4p}{(2\pi)^4}\bar{\Phi}_{P'}(k)
d_1(P'-k)\delta_2(k)K^{\mu}(P',k;P,p)d_1(P-p)
\delta_2(p)\Phi_{P}(p) .   \nonumber
\eea
Just as \eq{Mand} can be derived by gauging the BS equation for the two-body
Green function~\cite{G4d}, one can similarly show that \eq{Mand_x} results from
the gauging of the spectator equation for the two-body Green function, with
$\delta^\mu(p',p)$ being the result of gauging $\delta(p)$. Thus \eq{delta^mu}
(or \eq{new}) constitutes our answer to the question of how to gauge the
on-mass-shell particle propagator.

\section{Properties of the gauged on-mass-shell propagator}

\subsection{Gauge invariance}

In order to prove that the bound state current of \eq{Mand_x} satisfies current
conservation, all we need to do is follow the corresponding proof for the bound
state current of \eq{Mand} in the original four-dimensional BS approach of
Ref.~\cite{G4d}.  Indeed, to keep the correspondence with the four-dimensional
BS approach, we will use \eq{delta} for the propagator $\delta(p)$ and
\eq{delta^mu} for the gauged propagator $\delta^{\mu}(p',p)$, but we will {\em
not} get rid of the relative energy integration in \eq{Mand_x} (with the help
of the $\delta$-functions contained in $\delta$ and $\delta^\mu$). Thus our
derivation will look identical to the one in the four-dimensional BS approach,
except that particle 2 will have the propagator $\delta(p)$ instead of the usual
one $d(p)$.

Following this strategy, there is no need to repeat the proof of current
conservation here, except to note that a necessary ingredient in the proof of
the BS case is the Ward-Takahashi identity for the propagator $d(p)$. Thus to
prove current conservation for the three-dimensional expression of \eq{Mand_x},
it is sufficient to show that the on-mass-shell propagator $\delta(p)$ likewise
satisfies the Ward-Takahashi identity
\be
(p'_\mu-p_\mu)\delta^\mu(p',p)= ie \left[\delta(p)-\delta(p')\right] .\eqn{wti}
\ee

To prove \eq{wti}, all that is required is a simple evaluation of
$\delta^\mu(p',p)$ as given by \eq{delta^mu}. In the case of a spinor particle,
$\Gamma^\mu=e\gamma^\mu$, and one part of \eq{delta^mu} gives
\bea
(p'_{\mu}-p_{\mu})\delta(p')\Gamma^{\mu}(p',p)\Lambda(p)&=&
2\pi ie\delta^+ (p'^2-m^2)(\pslash\,'+m)(\pslash\,'-\pslash)(\pslash+m) \nn
&=&2\pi ie\delta^+ (p'^2-m^2)\left[(p'^2-m^2)(\pslash+m)
-(\pslash\,'+m)(p^2-m^2)\right]\nn
&=&-ie\delta(p')(p^2-p'^2).
\eea
Similarly the other part of \eq{delta^mu} gives
\be
(p'_{\mu}-p_{\mu})\Lambda(p')\Gamma^{\mu}(p',p)\delta(p)
=-ie\delta(p)(p^2-p'^2).
\ee
The Ward-Takahashi identity of \eq{wti} follows immediately. In the case of a
scalar particle the algebra showing \eq{wti} is even simpler.

\subsection{Charge conservation} 

Although current conservation, according Noether's theorem implies charge
conservation, in the currently used terminology ``charge conservation'' means
that the expression for the bound state electromagnetic current of \eq{Mand}
should give the charge of the composite system at zero momentum transfer if one
uses the relativistic normalization condition for the bound state vertex
function \cite{Yaz,Adam}. Here we shall show that this is the case also for the
three-dimensional expression of \eq{Mand_x} if one uses our choice for the
gauged on-mass-shell propagator, \eq{delta^mu}.

It is convenient at this stage to introduce a symbolic notation for some of our
equations. For example, we write the bound state BS equation, \eq{Phi-BS},
symbolically as
\be
\Phi_P=KG_0\Phi_P
\ee
where $G_0=d_1d_2$, and the corresponding equation for the bound state current,
\eq{Mand}, as
\be
j^\mu(P',P)=\la P'|J^\mu(0)|P\ra = \bPhi_{P'}(G_0^\mu+G_0K^\mu G_0)\Phi_P.
\eqn{Mand_sym}
\ee
Here we have also used the fact that the gauged two-particle propagator is given
by \cite{G4d}
\be
G_0^\mu = (d_1d_2)^\mu = d_1^\mu d_2 + d_1 d_2^\mu.
\ee
The spectator version of the above three equations is obtained by making the
replacement $G_0\rightarrow {\cal G}_0$ (which implies that $d_2\rightarrow
\delta_2$).  Below we shall occasionally use such symbolic notation without
further explanation.

To prove charge conservation for the three-dimensional spectator approach, we
use the philosophy outlined above; namely, we follow the proof of the
four-dimensional BS case only replacing the Feynman propagator of particle 2 by
our on-mass-shell version. The proof of the BS case relies on the fact that the
Feynman propagator $d(p)$ satisfies the Ward identity
\be
d^{\mu}(p,p)=-ie\frac{\partial d(p)}{\partial p_\mu} \eqn{ward_d}.
\ee
Similarly, because the interaction current $K^\mu$ is an input to the expression
of \eq{Mand_sym}, it too must be constructed to satisfy the Ward identity, which
in the two-particle case reads
\be
K^{\mu}(P,k;P,p)=-i\left[
e_2\frac{\partial K(P,k,p)}{\partial k_{\mu}}
+\frac{\partial K(P,k,p)}{\partial p_{\mu}}e_2
+(e_1+e_2)\frac{\partial K(P,k,p)}{\partial P_{\mu}}\right].
\ee
Combining the last two equations with the relativistic normalization condition
for the bound state vertex function:
\be
-i\bar{\Phi}_P\left(\frac{\partial G_0}{\partial P_{\mu}}+
G_0\frac{\partial K}{\partial P_{\mu}}G_0\right)
\Phi_P=2P_{\mu},                        \eqn{norm}
\ee
one then obtains the charge conservation condition
\be
\la P|J^{\mu}(0)|P\ra =
\bPhi_P\left(G_0^\mu+G_0K^\mu G_0\right)\Phi_P  \eqn{charge}
= 2Q P_{\mu}
\ee
where $Q$ is the total charge of the two-body system.
 
To show charge conservation in the three-dimensional case, we see that it is
sufficient to prove the Ward identity for our on-mass-shell propagator: 
\be
\delta^{\mu}(p,p)=-ie\frac{\partial\delta(p)}{\partial p_\mu}. \eqn{ward} 
\ee
The rest of the proof is the same as above, but with ${\cal G}_0$ everywhere
replacing $G_0$.

We shall prove \eq{ward} for the spinor particle case by again using a direct
evaluation of our expression for $\delta^\mu(p,p)$:
\bea
\lefteqn{\delta^{\mu}(p,p)=-2\pi i\Lambda(p)\gamma^{\mu}\Lambda(p)\frac
{\partial\delta^+(p^2-m^2)}{\partial p^2}}\nn
&&=-2\pi ie(\pslash+m)\gamma^{\mu}(\pslash+m)\frac
{\partial\delta^+(p^2-m^2)}{\partial p^2}\nn
&&=-2\pi ie\left[2p^{\mu}(\pslash+m)-\gamma^{\mu}(p^2-m^2)\right]\frac
{\partial\delta^+(p^2-m^2)}{\partial p^2}\nn
&&=-2\pi ie \left[(\pslash+m)\frac
{\partial\delta^+(p^2-m^2)}{\partial p_\mu}-\gamma^{\mu}\frac
{\partial(p^2-m^2)\delta^+(p^2-m^2)}{\partial p^2}+\gamma^{\mu}\frac
{\partial(p^2-m^2)}{\partial p^2}\delta^+(p^2-m^2)\right]\nn
&&=-2\pi ie\left[(\pslash+m)\frac
{\partial\delta^+(p^2-m^2)}{\partial p_\mu}+
\gamma^{\mu}\delta^+(p^2-m^2)\right]=-2\pi ie \frac
{\partial(\pslash+m)\delta^+(p^2-m^2)}{\partial p_\mu}\nn
&&=-ie\frac{\partial\delta(p)}{\partial p_\mu}.\nonumber
\eea
It should be emphasised that we did not try to obtain the Ward identity of
\eq{ward} from the Ward-Takahashi (WT) identity of \eq{wti} as there is an
ambiguity in extracting the value of $\delta^\mu(p,p)$ in this way.  Indeed a
good example of this ambiguity is the GR expression for the gauged on-mass-shell
propagator, \eq{dGR}, which satisfies the WT identity of \eq{wti} as well, but
does not satisfy the Ward identity of \eq{ward} since it differs from
\eq{delta^mu} by the term
\bea
\delta^{\mu}(p',p)-\delta'^{\mu}(p',p)
&\!\!=\!\!&-\delta(p')\Gamma^{\mu}(p',p)\delta(p)\nn
&\!\!=\!\!&-4\pi^2 \Lambda(p')\Gamma^{\mu}(p',p)\Lambda(p)
\delta^+(p'^2-m^2)\delta^+(p^2-m^2) \eqn{diff}
\eea
which does not vanish at zero momentum transfer. \eq{diff} is derived by
paying careful attention to the $i\epsilon$ terms present in the one-particle
propagators in $\delta'^\mu(p',p)$, and using the fact that for $q^2<4m^2$
\ben
\delta^-(p'^{2}-m^2)\delta^+(p^2-m^2)=\delta^+(p'^2-m^2)\delta^-(p^2-m^2)=0.
\een
This means that the use of \eq{dGR} does not lead to charge conservation (in
contrast to what is claimed in Ref.\ \cite{Adam}). The ambiguity of extracting
$\delta^\mu$ from the WT identity can be seen explicitly from the fact that
\bea
q_\mu\left[\delta^{\mu}(p',p)-\delta'^{\mu}(p',p)\right] &=&
-4\pi^2 (p'_\mu-p_\mu)\Lambda(p')\Gamma^{\mu}(p',p)\Lambda(p)
\delta^+(p'^2-m^2)\delta^+(p^2-m^2)\nn
&=&-4\pi^2(\pslash\,'+m)(\pslash\,'-\pslash)(\pslash+m)
\delta^+(p'^2-m^2)\delta^+(p^2-m^2)\nn
&=&0, \nonumber
\eea
while
\bea
\delta^{\mu}(p,p)-\delta'^{\mu}(p,p) &\ne& 0. \hspace{9.2cm}\nonumber
\eea
\subsection{Comparison of the two prescriptions}

In the previous discussion of charge conservation we found a significant
difference between our prescription for the gauged on-mass-shell propagator and
the one of GR at the point $q=0$. Here we would like to compare the two
prescriptions also for $q\ne 0$.

The first thing to note is that there is no difference between the two
prescriptions for $q^2>0$ as well as for $q^2=0$ (but $q\ne 0$), since the
product of the two $\delta$-functions in \eq{diff} will always be zero under
these conditions. Thus our prescription will not change the results of
Ref.~\cite{Sur} where pion photoproduction off a nucleon was calculated using
the GR prescription.  On the other hand, for $q^2<0$, which includes the case of
electron scattering, the contribution of \eq{diff} is not zero. We would
therefore like to investigate this difference between the two prescriptions when
applied to the two-body bound state current in the case where $q^2<0$. Writing
the bound state current symbolically as in \eq{Mand_sym}, the difference in
using the two prescriptions in \eq{Mand_sym} is clearly given by
\be
\Delta j^\mu(P',P)
= \bPhi_{P'} \left({\cal G'}_0^\mu - {\cal G}_0^\mu\right) \Phi_P.
\ee
Using \eq{diff}, numerically we have that
\bea
\lefteqn{\Delta j^\mu(P',P)
 =\int \frac{d^4p}{(2\pi)^4}\bar{\Phi}_{P'}(p')d_1(P-p)\delta_2(p')
\Gamma_2^{\mu}(p',p)\delta_2(p)\Phi_{P}(p)}\nn
&& =\int \frac{d^3p}{(2\pi)^2}\frac{1}{2\sqrt{\bfp^2+m^2}}
\bar{\Phi}_{P'}(p')d_1(P-p)\Lambda_2(p')\delta^+(p'^2-m^2)
\Gamma_2^{\mu}(p',p)\Lambda_2(p)\Phi_{P}(p)\nn
&& =\int \frac{d^3p}{(2\pi)^2}\frac{1}{2\sqrt{\bfp^2+m^2}}
\bar{\Psi}_{P'}(\bfp')\delta^+(p'^2-m^2)
\Gamma_2^{\mu}(p',p)d_1^{-1}(P-p)\Psi_{P}(\bfp)\nn
\eea
where we have introduced the wave function $\Psi_P({\bf p})$ defined by
\be
\Psi_P({\bf p})=d_1(P-p)\Lambda_2(p)\Phi_P({\bf p)}|_{p^0=\sqrt{{\bf p}^2+m^2}}.
\ee
For the scalar particle case in the Breit reference frame where $q_0=0$ and
${\bf P}'=-{\bf P}=\bfq/2$, we have that
\bea
\lefteqn{\Delta j^{\mu}(P',P)
= -i\int \frac{d^3p}{(2\pi)^2}\frac{1}{2\sqrt{{\bf p}^2+m^2}} 
\bar{\Psi}_{P'}(\bfp+\bfq)\Gamma_2^{\mu}(p',p)\Psi_{P}(\bfp) }\hspace{3cm}\nn
&&|{\bf q}|^{-1}\delta(2p_z+|\bfq|)
\left(M^2-\sqrt{{\bf q}^2+4M^2}\sqrt{{\bf p}^2+m^2}+\frac{{\bf q}^2}{2}
\right)       ,   \eqn{dj^mu}
\eea
where we have chosen the $z$-axis along $\bfq$ to write
\be
\delta(2\bfp\cdot\bfq+\bfq^2)=|{\bf q}|^{-1}\delta(2p_z+|\bfq|)
\ee
which is valid for $\bfq\neq 0$. From \eq{dj^mu} it is clearly seen that $\Delta
j^{\mu}(P',P)$ diverges as $\bfq\rightarrow 0$. 

To estimate the significance of $\Delta j^{\mu}(P',P)$ at values of $\bfq$ away
from zero, we may compare \eq{dj^mu} with the second term on the r.h.s. of
\eq{Mand_x}, which describes the contribution to the bound state current of
particle 1's gauged (Feynman) propagator:
\be
\int \frac{d^3p}{(2\pi)^2}\frac{1}{2\sqrt{{\bf p}^2+m^2}}
\bar{\Psi}_{P'}(\bfp)\Gamma_1^{\mu}(P'-p,P-p)\Psi_{P}(\bfp).
\ee
It can be seen that these two contributions are roughly of comparable size.

\section{Derivation}

Having established the validity of our expression of \eq{delta^mu} for the
gauged on-mass-shell particle propagator, in this Section we would like present
two ``derivations'' of this expression that can give a better insight into the
origin of this particular form.

\subsection{Connection with the four-dimensional approach}

Here we show that our gauged on-mass-shell particle propagator corresponds to
the contribution of the positive energy propagator poles of the corresponding
term in the four-dimensional BS expression for the bound state current.

The relevant term is the first term on the r.h.s. of \eq{Mand}:

\ben
A = \int \frac{d^4p}{(2\pi)^4} \bar{\Phi}_{P'}(p')d_1(P-p)
d_2(p')\Gamma_2^{\mu}(p',p) d_2(p)\Phi_{P}(p).
\een
Ignoring all poles in the complex $p_0$-plane except those contained in the
two $d_2$ propagators, we may close the $p_0$ integration contour in the bottom
half plane to obtain that
\bea
A &=&  -\int \frac{d^4p}{(2\pi)^4}
\frac{\bar{\Phi}_{P'}(p')d_1(P-p) \Lambda_2(p')
\Gamma_2^{\mu}(p',p) \Lambda_2(p)\Phi_{P}(p)}
{[(p_0+q_0)^2-\omega'^2+i\epsilon](p_0^2-\omega^2+i\epsilon)}\nn[3mm]
&\approx& 2\pi i \int \frac{d^4p}{(2\pi)^4}
\bar{\Phi}_{P'}(p')d_1(P-p)\Lambda_2(p')
\left[\frac{\delta_2^+(p'^2-m^2)}{p^2-m^2 \pm i\epsilon}+
\frac{\delta_2^+(p^2-m^2)}{p'^2-m^2 \mp i\epsilon}\right]
\Gamma_2^{\mu}(p',p) \Lambda_2(p)\Phi_{P}(p)\nn[2mm]
&=&  \int \frac{d^4p}{(2\pi)^4}
\bar{\Phi}_{P'}(p')d_1(P-p)
\left[\delta_2(p')\Gamma_2^{\mu}(p',p)d_2(p)
+d_2^{-}(p')\Gamma_2^{\mu}(p',p)\delta_2(p)\right]\Phi_{P}(p) , \eqn{pm}\\
&=&  \int \frac{d^4p}{(2\pi)^4}
\bar{\Phi}_{P'}(p')d_1(P-p)
\left[\delta_2(p')\Gamma_2^{\mu}(p',p)d^{-}_2(p)
+d_2(p')\Gamma_2^{\mu}(p',p)\delta_2(p)\right]\Phi_{P}(p),   \eqn{mp}
\eea
where it is important to notice that 
\be
d^-(p)=\frac{i\Lambda(p)}{p^2-m^2-i\epsilon}=d(p)-2\pi \Lambda(p)\delta(p^2-m^2)
\eqn{18}
\ee 
differs from the Feynman propagator $d(p)$ in the sign of the $i\epsilon$ term.
We can use either of the forms \eq{pm} or \eq{mp} to extract the gauged
on-mass-shell propagator since they both give the same result.
We can choose for example
\bea
\delta^\mu(p',p) &=& \delta(p')\Gamma^{\mu}(p',p)d(p)
+d^{-}(p')\Gamma^{\mu}(p',p)\delta(p)     \eqn{delta^mu-pm} \\[3mm]
&=&i\frac
{\delta(p')\Gamma^{\mu}(p',p)\Lambda(p)-
\Lambda(p')\Gamma^{\mu}(p',p)\delta(p)}{p^2-p'^2+i\epsilon}.
\eea
Noticing that the latter expression is regular at $p^2-p'^2=0$, it becomes clear
that the $i\epsilon$ term may be dropped from the denominator, in this way
giving our expression of \eq{delta^mu}.

Note that \eq{delta^mu-pm} is particularly useful for a comparison with the
prescription of GR given by \eq{dGR}. The difference lies in the sign of the
$i\epsilon$ term in $d^{-}(p')$. As shown above, this difference is crucial for
charge conservation.

\subsection{Derivation by minimal substitution}

It is well known that gauging a momentum dependent quantity by minimal
substitution $p^{\mu}\rightarrow p^{\mu}+eA^{\mu}(x)$ guarantees not only gauge
invariance but charge conservation as well. For this reason it would be
interesting to see if we can derive our form for $\delta^\mu(p',p)$ by
implementing the minimal substitution procedure in the on-mass shell propagator
$\delta(p)$. The way that this can be done is by expressing $\delta(p)$ in
terms of the difference of Feynman propagators:
\be
\Delta(p) \equiv d(p)-d^-(p)= \frac{i\Lambda(p)}{p^2-m^2+i\epsilon}-
\frac{i\Lambda(p)}{p^2-m^2-i\epsilon} =
2\pi\Lambda(p)\delta(p^2-m^2). \eqn{Delta} 
\ee
Thus $\delta(p) = \theta(p_0)\Delta(p)$. Now by implementing minimal
substitution in \eq{Delta} we will clearly obtain that
\bea
\Delta^\mu(p',p)
&=& d(p')\Gamma^{\mu}(p',p)d(p)-d^-(p')\Gamma^{\mu}(p',p)d^-(p)\nn
&=& \left[d(p')-d^-(p')\right]\Gamma^{\mu}(p',p)d(p)
+d^-(p')\Gamma^{\mu}(p',p)\left[d(p)-d^-(p)\right]\nn
&=&\Delta(p')\Gamma^\mu(p',p)d(p)+d^-(p')\Gamma^\mu(p',p)\Delta(p).
\eea
If we now drop the negative energy $\delta$-functions in the $\Delta$'s, we
derive the expression for the gauged on-mass-shell propagator
\be
\delta^\mu(p',p)
=\delta(p')\Gamma^\mu(p',p)d(p)+d^-(p')\Gamma^\mu(p',p)\delta(p)
\ee
which is the same result as \eq{delta^mu-pm}.

\section{Application to Deuteron Photodisintegration}

With the gauged on mass shell propagator specified, we now have all that is
needed to derive gauge invariant three-dimensional expressions within the
spectator approach for any system of hadrons interacting with an external
electromagnetic field. Here we would like to demonstrate our gauging procedure
by calculating the amplitude for deuteron photodisintegration.

As the hadronic system of interest here consists of two identical nucleons, some
of the previous expressions given for the distinguishable particle case need to
be slightly modified. In particular, the bound state spectator equation for
identical nucleons is given by  ($\Phi\equiv\Phi_P$)
\be
\Phi = \frac{1}{2}K {\cal G}_0 \Phi    \eqn{Phi-Gross-id}
\ee
where the kernel $K$ is the sum of all possible irreducible diagrams for
identical particles, and is therefore antisymmetric under the exchange of
nucleon labels.

In the four-dimensional approach of Ref.~\cite{G4d}, the $d\rightarrow N\!N$
transition current $j^\mu_0$ is given by
\be
j^\mu_0=G_0^{-1}[G_0\Phi]^\mu=\Phi^\mu+G_0^{-1}G^\mu_0\Phi    \eqn{j^mu_0-def}
\ee
where $\Phi^\mu$ is the gauged vertex function to be discussed shortly. The last
equality in \eq{j^mu_0-def} was obtained by using the rule for gauging products
\cite{G4d}.  To turn this BS expression into a three-dimensional one using the
spectator approach, all we need to do is replace the BS version of $\Phi$ by the
one that satisfies the spectator equation, \eq{Phi-Gross-id}. But we do not
replace $G_0$ by ${\cal G}_0$ in \eq{j^mu_0-def} as this would introduce an
unphysical $\delta$-function behaviour into the photoproduction amplitude. To
obtain an expression for $\Phi^\mu$ we gauge \eq{Phi-Gross-id}:
\be
\Phi^\mu=\frac{1}{2}\left(K^\mu{\cal G}_0\Phi
+K{\cal G}_0^\mu\Phi+K{\cal G}_0\Phi^\mu\right)
\ee
which may be solved for $\Phi^\mu$ giving
\be
\Phi^\mu=\frac{1}{2}\left(1-\frac{1}{2}K{\cal G}_0\right)^{-1}
\left(K^\mu{\cal G}_0\Phi+K{\cal G}_0^\mu\Phi\right).      \eqn{Phi^mu-temp}
\ee
To simplify this expression we use the equations for the two-nucleon t-matrix
$T$. In the spectator approximation they are given by
\be
T= K + \frac{1}{2}K{\cal G}_0 T = K + \frac{1}{2}T{\cal G}_0 K,  \eqn{T}
\ee
from which the relation
\be
\left(1-\frac{1}{2}K{\cal G}_0\right)^{-1}=1 + \frac{1}{2}T{\cal G}_0
\ee
follows. Using this in \eq{Phi^mu-temp} we obtain that
\be
\Phi^\mu=\frac{1}{2}\left(1+\frac{1}{2}T{\cal G}_0\right)
K^\mu{\cal G}_0\Phi 
+
\frac{1}{2}T{\cal G}_0^\mu\Phi 
\ee
where
\be
{\cal G}_0^\mu=(d_1\delta_2)^\mu=d_1^\mu\delta_2+d_1\delta_2^\mu.
\ee
The $d\rightarrow N\!N$ transition current is therefore given by
\be
j^\mu_0 = \left[\Gamma_1^\mu d_1 + \Gamma_2^\mu d_2
+ \frac{1}{2}\left(1+\frac{1}{2}Td_1\delta_2\right)K^\mu d_1\delta_2
+ \frac{1}{2}T \left(d_1\delta_2^\mu + d_1^\mu\delta_2\right)\right]\Phi 
\eqn{j^mu_0}
\ee
where we have used \eq{d^mu}. This expression can be used to calculate the
deuteron photodisintegration amplitude by contracting \eq{j^mu_0} with the
photon polarisation vector $\varepsilon_\mu$.. 

An interesting aspect of \eq{j^mu_0} is the appearance of the Feynman propagator
$d_2$ in the second term on the r.h.s., while in all other parts of the equation
(including the equation for $T$, \eq{T}) the on-mass-shell propagator $\delta_2$
is used. This is of course a consequence of us having used $G_0$ instead of
${\cal G}_0$ in \eq{j^mu_0-def}. Using $d_2$ here is reasonable since it is not
inconsistent with the spectator approach, and it avoids the unphysical behaviour
of amplitudes that would result if $\delta_2$ were used instead. On the other
hand, it is not entirely clear if this singular use of $d_2$ will affect the
gauge invariance of the electromagnetic transition current $j^\mu_0$. We shall
therefore show explicitly that the expression for $j^\mu_0$ given by
\eq{j^mu_0-def} does indeed satisfy gauge invariance despite the use of $G_0$ in
this equation.

\eq{j^mu_0-def} is a symbolic equation whose numerical form simplifies down to
\bea
j^\mu_0(k_1,k_2;P) = \Phi_P^\mu(k_1,k_2)
&+& d^{-1}(k_1)d^\mu(k_1,k_1-q)\Phi_P(k_1-q,k_2)\nn
&+& d^{-1}(k_2)d^\mu(k_2,k_2-q)\Phi_P(k_1,k_2-q)    \eqn{j^mu_0-ex}
\eea
where we show the momenta of both particles explicitly, and it is understood 
that $k_1+k_2=P+q$ where $q$ is the momentum of the incoming photon.
By construction, the input quantities $K^\mu$, $d^\mu$, and $\delta^\mu$
satisfy the WT identities
\bea
-iq_\mu K^\mu(p'_1p'_2;p_1p_2)
&=& e_1K(p'_1-q,p'_2;p_1p_2)-K(p'_1p'_2;p_1+q,p_2)e_1\nn
&+& e_2K(p'_1,p'_2-q;p_1p_2)-K(p'_1p'_2;p_1,p_2+q)e_2, \eqn{wti-K} \\[3mm]   
-iq_\mu d^\mu(p',p) &=& ed(p)-d(p')e,  \eqn{wti-dd}\\[3mm]
-iq_\mu \delta^\mu(p',p) &=& e\delta(p)-\delta(p')e, \eqn{wti-ddelta}
\eea
respectively. In \eq{wti-K} we again use a notation where the momentum of each
particle is shown explicitly, and where ${p'_1+p'_2=p_1+p_2+q}$. In \eqs{wti-dd}
and (\ref{wti-ddelta}) we similarly have that $p'=p+q$. Using these relations
it is easy to show that the WT identity for $\Phi^\mu$ is given by
\be
-iq_\mu \Phi_P^\mu(k_1,k_2) = e_1\Phi_P(k_1-q,k_2)+e_2\Phi_P(k_1,k_2-q),
\eqn{wti-Phi}
\ee
where $k_1+k_2=P+q$. Then using the WT identities for $d^\mu$ and $\Phi^\mu$
in calculating the divergence of \eq{j^mu_0-ex}, we obtain that
\be
q_\mu j^\mu_0(k_1,k_2;P)=ie_1d^{-1}(k_1)d(k_1-q)\Phi_P(k_1-q,k_2)+
ie_2d^{-1}(k_2)d(k_2-q)\Phi_P(k_1,k_2-q)
\ee
which is zero for on-mass-shell nucleons ($k_1^2=k_2^2=m^2$).

\section{Summary}

In this work we have shown how to construct three-dimensional integral equations
that describe a system of hadrons and their interaction with an external
electromagnetic field. The equations are relativistic (covariant), unitary,
gauge invariant, and conserve charge. Our method is based upon a recent work
where we show how four-dimensional integral equations of quantum field theory
can be gauged so that an external photon is coupled to all possible places in
the underlying strong interaction perturbation graphs, without the need to do a
perturbation expansion \cite{G4d}.

The starting point of our construction is a set of four-dimensional integral
equations of relativistic quantum field theory describing the system of hadrons
in questions. For example, for the two-nucleon system below pion production
threshold the starting point would be the Bethe-Salpeter equation, while above
pion production threshold the equations of Ref.~\cite{4d} would be appropriate.
We do not gauge these equations at this stage, but instead convert them to the
spectator equations of Gross \cite{Gross} by the introduction of the ``on mass
shell propagator'' $\delta$. The modified four-dimensional equations are then
gauged just in the same way as was done for the four-dimensional equation of
field theory.

The three-dimensional reduction then rests on the construction of a gauged
on-mass-shell propagator $\delta^\mu$. A $\delta^\mu$ that satisfies both the
Ward-Takahashi and Ward identities is necessary for the gauge invariance and
charge conservation of the final equations. We have shown how such a gauged
on-mass-shell propagator can be constructed, and compared our results with what
was proposed in the literature \cite{GR}.  With $\delta^\mu$ specified, we then
demonstrated our gauging procedure by constructing the amplitude for deuteron
photodisintegration within the spectator approach.

Our gauging procedure can be easily applied to more complicated systems. For
example, in Ref.~\cite{nnn3d} we have used it to derive gauge invariant
three-dimensional expressions for the gauged three-nucleon system. It also does
not depend on the nature of the external gauge field. Thus it can equally well
be used to describe the weak interactions of hadronic systems.

\acknowledgments

The authors would like to thank the Australian Research Council for their
financial support.

\appendix
\section*{}

In the above discussion our particles were assumed to be structureless. Here we
show one way to include electromagnetic form factors that preserves gauge
invariance and charge conservation. Our approach is close in spirit to the one
used by Gross and Riska \cite{GR}.

As before, the three-dimensional reduction is effected by the replacement
\be
d(p)=\frac{i\Lambda(p)}{p^2-m^2+i\epsilon} \hspace{5mm}\rightarrow\hspace{5mm}
\delta(p)= 2\pi \Lambda(p) \delta^+(p^2-m^2) ,
\ee
but where now (for spinor particles) $\Lambda(p)\neq \not\!p+m$ because of
dressing included in $d(p)$. Nevertheless, the on-mass-shell particle propagator
$\delta(p)$ is not affected by dressing except for an overall renormalization
constant $Z$:
\be
\delta(p)=2\pi \Lambda(p)\delta^+(p^2-m^2)=2\pi Z(\pslash+m)
\delta^+(p^2-m^2) .       \eqn{delta-Z}
\ee
The latter result follows from the spectral decomposition of the dressed
Feynman propagator
\be
d(p)=\frac{Z(\pslash+m)}{p^2-m^2+i\epsilon}+R(p)\eqn{sr}
\ee
where $R(p)$ is a function that is regular at $p^2=m^2$.
Although the on-mass-shell particle propagator of \eq{delta-Z} cannot have
dressing in the usual sense, one can nevertheless introduce an electromagnetic
form factor into the gauged on-mass-shell propagator through the definition
\be
\delta^{\mu}(p',p)=i\frac
{\delta(p')\Gamma^{\mu}(p',p)\Lambda(p)-
\Lambda(p')\Gamma^{\mu}(p',p)\delta(p)}{p^2-p'^{2}}
\ee
or in the explicitly regular form 
\be
\delta^{\mu}(p',p)=2\pi i\Lambda(p')\Gamma^{\mu}(p',p)\Lambda(p)\frac
{\delta^+(p'^{2}-m^2)-\delta^+(p^2-m^2)}{p^2-p'^{2}}   \eqn{new-Z}
\ee
where now the electromagnetic vertex function $\Gamma^{\mu}(p',p)\ne
e\gamma^\mu$, but does satisfy the Ward-Takahashi identity
\be
(p'-p)_{\mu}\Gamma^{\mu}(p',p)
=ie\left[d^{-1}(p')-d^{-1}(p)\right].    \eqn{wti-Gamma}
\ee
On the mass shell $\Gamma^{\mu}$ takes on the usual form
\be
\Gamma^{\mu}(p',p)|_{p'^{2}=p^2=m^2}
=e\left[F_1(q^2)\gamma^{\mu}+i\frac{\sigma^{\mu\nu}}{2m}q_\nu F_2(q^2)\right].
\ee
Thus we have at the same time a structureless on-mass-shell propagator,
\eq{delta-Z}, together with a gauged on-mass-shell propagator that does have
structure, \eq{new-Z}. We will now show that together they nevertheless satisfy
the WT identity. This means that the structure described by \eq{new-Z} does not
contribute to the WT identity.

\subsection{Gauge invariance}

As in the structureless case, to prove gauge invariance of the theory it is
sufficient to show that the on-mass-shell propagator $\delta$ satisfies the WT
identity. We show this by explicitly evaluating $\delta^\mu$; however, as an
explicit form for $\Gamma^\mu$ is no longer available, we make use of
\eq{wti-Gamma} instead:
\bea
\lefteqn{(p'-p)_{\mu}\delta^{\mu}(p',p)= 
 -2\pi e\Lambda(p')\left[d^{-1}(p')-d^{-1}(p)\right]\Lambda(p)
\frac{\delta^+(p'^{2}-m^2)-\delta^+(p^2-m^2)}{p^2-p'^{2}} }\hspace{2cm}\nn
&& =2\pi e\Lambda(p')\left[d^{-1}(p')\Lambda(p)
\frac{\delta^+(p^2-m^2)}{m^2-p'^{2}} 
+d^{-1}(p)\Lambda(p)\frac{\delta^+(p'^2-m^2)}{p^2-m^{2}} \right]\nn
&& = 2\pi ie\left[\Lambda(p)\delta^+(p^2-m^2) 
- \Lambda(p')\delta^+(p'^2-m^2) \right]\nn
&& = ie\left[\delta(p)-\delta(p') \right]       \eqn{wti-delta}
\eea
where we used that
\be
d^{-1}(p')\delta^+(p'^{2}-m^2)=d^{-1}(p)\delta^+(p^{2}-m^2)=0.
\ee

\subsection{Charge conservation}

To prove charge conservation, it is again sufficient to show that the
on-mass-sell propagator $\delta$ satisfies the Ward identity. As in the
structureless case, the Ward identity cannot be deduced unambiguously from the
WT identity of \eq{wti-delta}, and therefore must be shown explicitly. This we
do by using \eq{new-Z} in the limit of zero momentum transfer:
\be
\delta^{\mu}(p,p)=-2\pi i\Lambda(p)\Gamma^{\mu}(p,p)\Lambda(p)\frac
{\partial \delta^+(p^2-m^2)}{\partial p^2} .
\ee 
To evaluate $\Gamma^{\mu}(p,p)$ we use its Ward identity,
\be
\Gamma^{\mu}(p,p)=ie\frac{\partial d^{-1}(p)}{\partial p_\mu}.
\ee
Then
\bea
\frac{1}{2\pi e}\delta^{\mu}(p,p)&=&\Lambda(p)\frac{\partial d^{-1}(p)}
{\partial p_\mu}\Lambda(p)\frac{\partial\delta^+(p^2-m^2)}{\partial p^2}\nn
&=&\frac{\partial}{\partial p^2}\left[\Lambda(p)\frac{\partial d^{-1}(p)}
{\partial p_\mu}\Lambda(p)\delta^+(p^2-m^2)\right]-
\frac{\partial}{\partial p^2}\left[\Lambda(p)\frac{\partial d^{-1}(p)}
{\partial p_\mu}\Lambda(p)\right]\delta^+(p^2-m^2).\nn   \eqn{temp}
\eea
Now
\bea
\Lambda(p)\frac{\partial d^{-1}(p)}{\partial p_\mu}\Lambda(p)&=&
\Lambda(p)\left[\frac{\partial d^{-1}(p)\Lambda(p)}{\partial p_\mu}-
d^{-1}(p)\frac{\partial \Lambda(p)}{\partial p_\mu}\right]\nn[1mm]
&=&-i\left[\Lambda(p)\frac{\partial (p^2-m^2)}{\partial p_\mu}-
(p^2-m^2)\frac{\partial \Lambda(p)}{\partial p_\mu}\right]\nn[2mm]
&=&-i\left[2p^\mu\Lambda(p)-
(p^2-m^2)\frac{\partial \Lambda(p)}{\partial p_\mu}\right]
\eea
so that
\be
\Lambda(p)\frac{\partial d^{-1}(p)}
{\partial p_\mu}\Lambda(p)\delta^+(p^2-m^2)=-2ip^\mu\Lambda(p)\delta^+(p^2-m^2).
\ee
This last relation is similar to the bound state normalization condition and
can also be deduced by taking the residues at $p^2=m^2$ in the identity
\ben
d(p)d^{-1}(p)d(p)=d(p).
\een

Substituting the last two results into \eq{temp}, we obtain
\bea
\frac{i}{2\pi e}\delta^{\mu}(p,p)
&=&\frac{\partial}{\partial p^2}\left[2p^\mu\Lambda(p)\delta^+(p^2-m^2)\right]-
\frac{\partial}{\partial p^2}\left[2p^\mu\Lambda(p)-
(p^2-m^2)\frac{\partial \Lambda(p)}{\partial p_\mu}\right]\delta^+(p^2-m^2)\nn 
&=&2p^\mu\frac{\partial}{\partial p^2}\left[\Lambda(p)\delta^+(p^2-m^2)\right] 
\eea
where we used that 
\ben
(p^2-m^2)\frac{\partial}{\partial p^2}\left[
\frac{\partial \Lambda(p)}{\partial p_\mu}\right]\delta^+(p^2-m^2)=0
\een
with all other terms cancelling. We thus find that
\be
\delta^{\mu}(p,p) 
= -2ip^\mu e \frac{\partial \delta(p)}{\partial p^2}
= -ie \frac{\partial \delta(p)}{\partial p_\mu}
\ee
as required.

% now the references. delete or change fake bibitem. delete next three
%   lines and directly read in your .bbl file if you use bibtex.

% \begin{figure}
% \caption{}
% \label{}
% \end{figure}

% \begin{table}
% \caption{}
% \label{}
% \end{tabular}
% \end{table}
\end{document}